# THE GRANDPARENT SCAM: A SYSTEMS PERSPECTIVE CASE STUDY ON ELDER FRAUD AND THE CONCEPT OF HUMAN LAYERING

Michelle Espinoza
Marymount University, Arlington, USA
M0e73021@marymount.edu

## Introduction

In April 2024, an 81-year-old Ohio man was charged with murder, assault, and kidnapping (Finley, 2024). The man believed that he was protecting his family from scammers threatening harm. What he did not realize was that the 61-year-old Uber driver he killed, was also a victim of the same scammers. Courier fraud is a form of fraud involving perpetrators posing as couriers or law enforcement officials to trick individuals into handing over money, personal information, or valuable items. Often, the couriers are unwitting victims unaware that they are part of a muling scam. The tragedy that occurred in Ohio also goes by another name, the Grandparent Scam (Federal Trade Commission, 2024). These fraudulent schemes frequently target vulnerable populations, such as the elderly or those less familiar with digital scams, making it a pressing issue that requires awareness and vigilance to reduce victimization.

This case study examines some common variants of the Grandparent Scam from a systems perspective and how weaponization of conscience is used in these scams. Additionally, this study examines the parallels between layering in money laundering and human layering in the execution of these scams. Cybercrime and online fraud are complex systems, curtailing their damage requires a holistic and interdisciplinary approach (M. D. Espinoza, 2024). This chapter is geared towards a multidisciplinary audience and assumes varying levels of expertise in the subtopics contained herein.

## Problem Statement

To address the growing crisis of elder fraud, the US Department of Justice formed the Transnational Elder Fraud Strike Force (US Department of Justice, 2019). The task force's purpose is to facilitate collaboration among state, local and federal law enforcement, and other federal agencies. Though recovery of stolen funds is rare, especially if the money is transferred overseas, the task force has been successful in coordinating investigation activities across multiple government agencies (White, 2022). Financial exploitation of the elderly has been linked to increased mortality, hospitalization rates, and poor physical and mental health—each of which add additional costs to the healthcare system that are difficult to measure in isolation, particularly since these crimes are underreported (Office for Victims of Crime, 2024). The low risk of prosecution and high profitability of elder fraud coupled with an aging population create a self-reinforcing positive feedback loop that if left unchecked, will result in the system's collapse. Understanding how and why fraudsters' modi operandi succeed is essential to constructing an effective defense (Rutherford, 2017).

**Method**

For the purpose of this research, an approach based on case study methodology was employed to examine the application of system dynamics within the context of the Grandparent Scam. This analysis involved the application of Milgram's classic experiment on obedience to authority, McAllister's model of trust encompassing two dimensions, and the concept of human layering. The specific case chosen for this investigation pertains to the tragic incident in 2024 involving the murder of a rideshare driver by an elderly individual, both of whom fell victim to the same group of scammers. In this scenario, the elderly man was targeted and exploited through the Grandparent Scam, while the rideshare driver experienced a dual victimization. Initially, she was deceived by the scammer who manipulated her into participating in a courier fraud scheme, and subsequently, she met a tragic fate at the hands of the elderly perpetrator.

The data utilized for this case study was sourced from a variety of outlets including official records, media reports, and academic publications. The information gathered was focused on gaining a comprehensive understanding of the specifics surrounding the chosen case, details of similar incidents, official documentation highlighting the prevalence of the scam, the cognitive processes influencing decision-making, and the ultimate repercussions stemming from the events. By delving into these various sources, a more nuanced and rich analysis of the manifestation of the weaponization of conscience in the Grandparent Scam within the context of the tragic 2024 killing of the rideshare driver by the elderly man was made possible. This comprehensive approach enables an exploration of the intricate dynamics at play in such instances of exploitation and victimization, shedding light on the complexities inherent in cases involving manipulation and criminal coercion. Ultimately, this research contributes to the broader understanding of the psychological, sociological, and ethical dimensions associated with the weaponization of conscience in fraudulent schemes and criminal activities.

**Systems and Multiple Feedback Loops**

Complexity science is the study of the relationship between systems. At its inception in the 1950s, system dynamics focused primarily on corporate and industrial problems such as production cycles, and supply chains. It relied heavily on hand-drawn simulations and calculations before computer modeling was available. Today, the principles of system dynamics are applied across various disciplines and industries including public policy, healthcare, and social systems (Gentili, 2021). The transferable principles of dynamic systems and amenability to qualitative manual model construction make system dynamics well-suited for modeling the weaponization of conscience. A system is a set of things working together as parts of a mechanism or interconnecting network. The conscience is a complex subsystem within a larger complex system. A feedback loop is a circular flow of information or effects that cause the system to regulate or adjust its behavior based on the information returned in each cycle (J. W. Forrester, 2003). Cycles can vary in length and systems can contain multiple feedback loops. The following example in Figure 1 illustrates how multiple agents in a system can seek the same goal and receive the same feedback loop, but each agent experiences variable time delays and makes different adjustments or alterations to their behavior based on the circular flow of information.

If a comedian tells a joke to an audience, each person may find varying levels of humor in the punchline. The comedian uses the joke to elicit laughter from multiple audience members (consciences). The varying degree of responses from each individual audience member interacts to produce the collective sound of laughter. Some individuals within the audience may laugh more intensely merely because of the contagion effect from hearing others' laughter. If the joke is successful, the comedian may add onto the joke in hopes of extending the audience's laughter. Conversely, if the audience does not laugh the comedian may switch to another topic. Laughter from the audience serves as the feedback loop that alters the comedian's behavior. The comedian is likely to alter their current and future routine based on the audience's level of laughter.

For much of the audience, the degree to which they enjoy the comedic routine influences their willingness to pay to see the comedian again in the future. Laughter is also the audience's feedback loop, but there is a longer delay in their behavior adaptation than that of the comedian. Some audience members may find the comedian's joke offensive and refrain from laughing. This is an example of a negative feedback loop that counters or could reduce the overall level of laughter from the audience. Unless enough people find the joke offensive and refrain from laughing, the negative feedback loop will not have a noticeable effect on the overall level of laughter from the audience. Conversely, the contagion effect may lead some audience members to laugh more simply because they see others laughing. The laughter contagion effect is an example of synergy in complex systems whereby the interaction of each agent's laughter combines to produce a more powerful effect or output than would be produced by any individual audience member.

Lastly, the comedy club owner needs to sell tickets for future events. Ticket sales serve as one feedback loop to indicate whether the comedy club owner should book this comedian in the future. The level of laughter from the audience also helps the club owner determine the likelihood of future ticket sales.

Every agent's decision is made in the context of a feedback loop (J. Forrester, 1968). Positive feedback loops are like the gas pedal in a vehicle and negative feedback loops are akin to its brakes. The brakes from a bicycle would be insufficient to stop a moving train. Similar to cybercrime, the negative feedback loops in the cybercrime system have been insufficient to slow its exponential growth. Learning to identify feedback loops is a quintessential skill in understanding system dynamics.

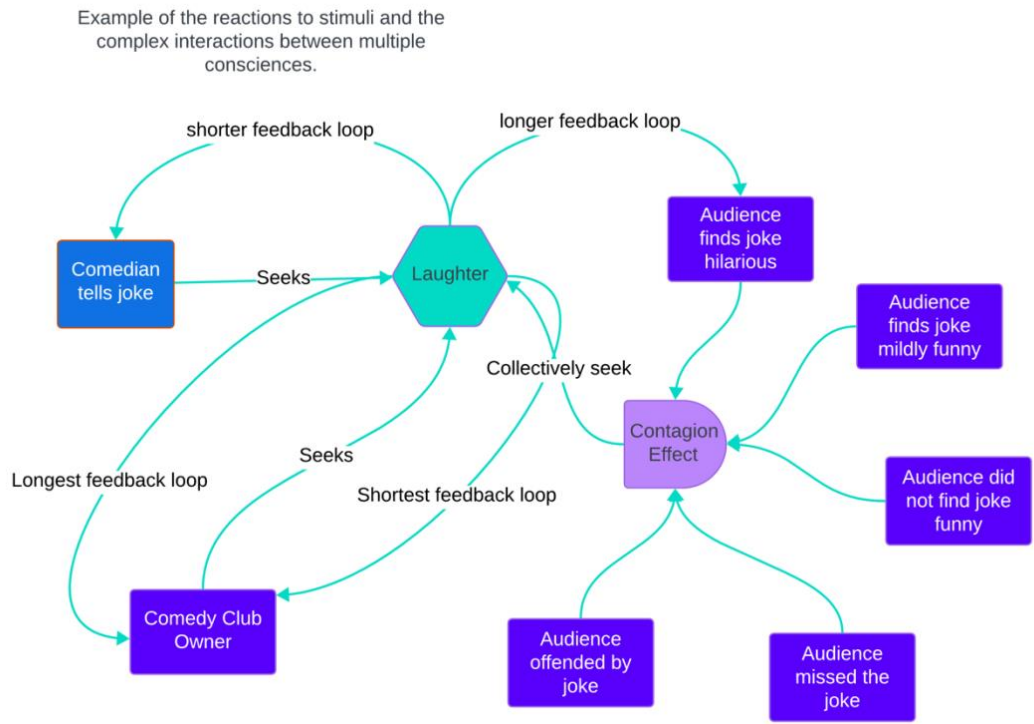

*Figure 1 Feedback Loop Example- Comedy. Note: Author's own work. Copyright attribution not required.*

Complex systems obey the laws of conservation and accumulation. The law of conservation states that energy cannot be created or destroyed and can only be transferred or transformed. This principle helps maintain the stability and balance of the system. On the other hand, accumulation refers to the process by which these systems gather or accumulate certain elements over time. This accumulation can lead to growth, development, or even disruptions within the system. Understanding how conservation and accumulation operate within complex systems is crucial for predicting their behavior and managing their dynamics effectively. Feedback loops and the laws of conservation and accumulation are used extensively in modeling the weaponization of conscience.

A key principle of information warfare is that human decision-making processes are the ultimate target for offensive information operations (Masakowski & Blatny, 2023; O'Brien, 2001). Weaponization of conscience is a tool/tactic employed by fraudsters to deceive their victims, camouflage their activity, and extend the effectiveness of their modi operandi (M. Espinoza, 2024). When conscience is modeled as a complex system, tweaking subcomponents of the decision system can predictably alter the outcome. The weaponization of conscience is a complex subsystem within the larger complex system of social engineering. For this framework, conscience is defined as an intrinsic motivational force guided by a person's moral beliefs. Those moral beliefs may or may not be rooted in religion and a person's moral beliefs can change or evolve. The fraudster's objective in the Grandparent Scam is to influence their victim's behavior and reasoning through information and emotions. Modeling the scam as a system enables practitioners and individuals to identify vulnerabilities and strengthen defenses against similar schemes in the future.

**Human Layering**

In the Ohio case, the scammer sent an unwitting victim to retrieve a package from another victim. This is where the cycle of victimization can take a dark turn, leading some victims to become unwitting perpetrators themselves. The sophisticated tactics employed by scammers often involve manipulating individuals into carrying out illegal activities under the guise of helping the scammers or avoiding legal consequences. This form of forced criminality preys on the vulnerabilities of victims, coercing them into committing acts they would never have considered otherwise, and the tactic expands the web of scammers making it increasingly difficult to distinguish victims from perpetrators.

Imagine for a moment that the shooter's weapon had been turned against him or that the Uber driver would have encountered a third scammer upon delivering the package and had been extorted into helping execute additional scams. This is a part of the dimension problem with cybercrime and online fraud. Scammers are not bound by geographic borders, and they seamlessly commit crimes between the physical and virtual world.

Scammers enlist co-conspirators through popular apps, they may launder funds through thousands of globally distributed mules. As the scam web expands through more jurisdictions, ensnares more victims, and as some of those victims become perpetrators themselves – a single scam can become so cost-prohibitive to investigate and prosecute that most law enforcement agencies will be unwilling to pursue it.

It is not only difficult to pinpoint the nexus of the crime, but it is also difficult to determine the jurisdiction in which the crime(s) should or can be prosecuted (Cross, 2020). In this context, the victims-turned-perpetrators are not only robbed of their money and dignity but also stripped of their agency and autonomy, becoming pawns in the scammers' schemes.

In the world of financial crimes, the parallel between money launderers layering money into the financial system and scammers concealing their identity by using mules as intermediaries reveals a striking similarity in their deceptive tactics and strategies to evade detection and scrutiny.

Money launderers employ a technique known as layering to disguise the origins of illegally obtained funds by moving them through a series of complex financial transactions (Matić Bošković, 2023). This process involves transferring money between multiple accounts and jurisdictions, making it difficult to trace the illicit funds back to their criminal source. By layering the money into the legitimate financial system, money launderers create a veil of legitimacy that obscures the illicit nature of the funds, allowing them to reintroduce the money into the economy without raising suspicion. Similarly, scammers utilize mules as intermediaries to conceal their true identities and distance themselves from their fraudulent activities (Chaganti et al., 2021). Mules are individuals who are recruited by scammers to transfer money or goods on their behalf, often unaware that they are participating in illegal schemes. By using mules as intermediaries, scammers can reduce their direct involvement in the illicit transactions, making it harder for law enforcement agencies to track down and apprehend the masterminds behind the scams. This tactic of using intermediaries allows scammers to operate from a safe distance, shielding themselves from accountability and prosecution. The parallel between money launderers layering funds and scammers using mules as intermediaries underscores the intricate web of deception and obfuscation employed by criminals in the financial realm. Both practices involve creating layers of complexity and intermediaries to obfuscate the trail of illicit activities, making it challenging for authorities to uncover and dismantle criminal networks. By understanding this parallel, law enforcement agencies and regulatory bodies can enhance their efforts to combat financial crimes and fraudulent schemes by targeting the layers of deception and intermediaries that facilitate illicit activities.

**The Grandparent Scam**

The attack surface in cybersecurity refers to the sum of all possible points where an unauthorized user could access or extract data from an organization's systems and networks. In the realm of cybercrime and online fraud, the attack surface exceeds 8 billion people.

At any specific juncture during the execution of a fraudulent scheme, the scammer may find themselves transitioning from the role of the perpetrator to that of the victim, or vice versa. It is plausible to encounter complex scenarios where deceitful activities are nested within one another, such as in situations in which a courier is leasing a compromised driver account obtained through a separate fraudulent operation (Skiba, 2021). The courier could be complicit in the illegal activity, coerced into participation because of extortion tactics, unknowingly utilized as a carrier of illicit goods, or tragically end up as a victim of a violent crime, mirroring the grim events detailed in the Ohio case analysis.

Multiple variants of the Grandparent Scam have been documented, the most common of which are depicted in Figure 2. The fraudulent scheme typically commences with a cunningly crafted phone call that preys on the victim's inherent desire to feel valued and indispensable. The scam artist then skillfully maneuvers to isolate the victim, employing tactics such as instilling fear by portraying scenarios where cooperation seems imperative for the protection of their family or loved ones. Alternatively, the scammer may leverage emotions of affection, persuading the victim that the grandchild is in dire need and that maintaining secrecy regarding the solicited assistance is crucial.

As the final stage of the scheme unfolds, the scammer brazenly requests for monetary aid, often devising elaborate plans for cash pick-up through a designated courier service. In instances where the targeted grandparent is familiar with online payment applications, the scammer may opt to facilitate the transaction through digital platforms, further complicating the fraudulent process.

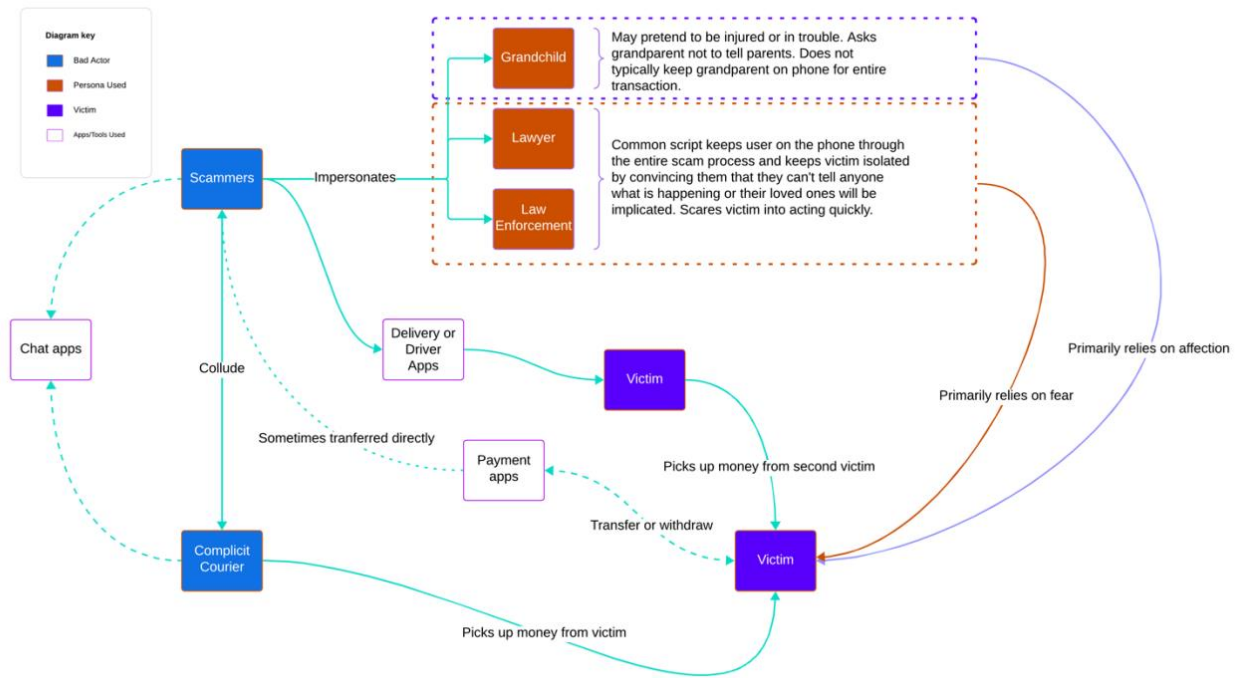

*Figure 2 Grandparent scam*

Given that a significant 83% of elderly individuals aged 65 and above in the United States acknowledge having grandchildren, fraudsters have a higher probability of success when targeting this demographic by pretending to be a distressed grandchild, leading the victim to react positively to the scammer's solicitations (Parker et al., 2015). The call recipient may or may not have a grandchild of the same gender or age range that the fraudster is impersonating. Perhaps to reduce the risk of failure, the fraudster poses as law enforcement and convinces the call recipient that they, the scammer, are the only one who can help the victim. To avoid negative feedback loops from friends or family of the victim that may hinder the fraudster's ability to reach their goal of cashing out, the fraudster may attempt to keep their victim on the phone until payment is secured or scare the victim into believing that if the victim reveals what is happening to anyone around them, those people may be implicated in a crime. If the scammer poses as the grandchild and relies on affective trust between the victim and their grandchild, perhaps the scammer can target more victims in the same timeframe that it would take for the scammer to pose as law enforcement and keep the victim on the phone until payment is secured. Posing as law enforcement may lower the scammer's risk of arrest.

According to McAllister (1995) there exist two dimensions of trust, cognitive and affective. When a scammer pretends to be a grandchild in a fraudulent scheme, they often exploit the emotional bond between the grandparent and grandchild to manipulate the victim into completing the scam. A visual representation of the conscience decision system can be observed in Figure 3. Dual-process theory of moral judgement asserts that moral decisions are the result of either one of two distinct mental processes, intuitive or conscious (Greene, 2009). Greene's dual-process model of moral judgement posits that intuitive judgements are fast, emotionally-driven, and primarily deontological (rule-based) whereas conscious decisions are deliberative and characteristically utilitarian (Greene, 2009). The cognitive mechanisms linked to affective trust tend to align with intuitive decision-making processes, whereas cognitive trust is more closely associated with deliberate and conscious decision-making processes. The distinction between these two types of trust sheds light on how individuals navigate moral dilemmas and make ethical choices. Thus, understanding the interplay between cognitive and affective trust can provide valuable insights into the complexities of human decision-making processes, particularly in situations involving deception and manipulation.

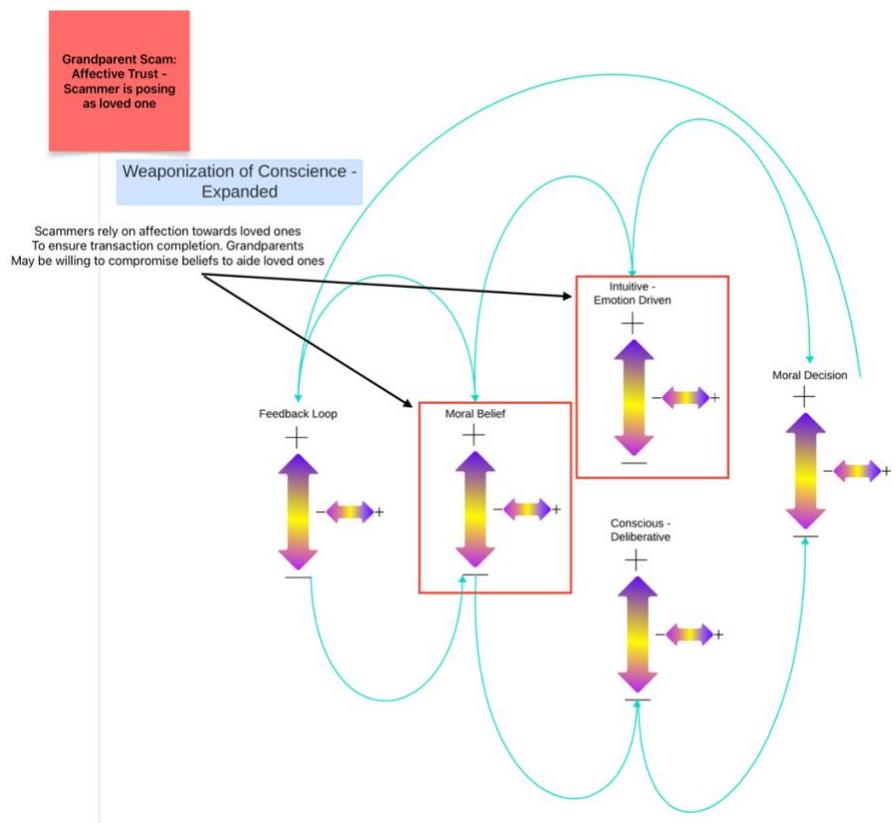

*Figure 3 Weaponization of Conscience*

**Milgram Experiment Obedience to Authority**
The Milgram experiment was a series of social psychology experiments conducted by Stanley Milgram in the 1960s. The primary aim of the experiment was to investigate the willingness of participants to obey an authority figure who instructed them to perform acts conflicting with their personal conscience. In the experiment, participants were told to administer electric shocks to another person (who was actually an actor pretending to be receiving the shocks) whenever they answered questions incorrectly. The shocks were fake, but the participants were unaware of this. The key finding of the Milgram experiment was that a surprisingly high percentage of participants were willing to administer what they believed to be potentially lethal electric shocks to the other person simply because they were instructed to do so by an authority figure. The primary outcome of the Milgram experiment was the demonstration of the powerful influence of authority on human behavior. It revealed the extent to which individuals are willing to obey authority figures, even when it involves acting against their own moral beliefs. The experiment raised ethical concerns about the treatment of participants but provided valuable insights into obedience, conformity, and the impact of authority figures on decision-making.

In the realm of fraudulent schemes, the tactic of a scammer posing as law enforcement draws intriguing parallels to the Milgram experiment and the concept of obedience to authority. Just as Stanley Milgram's social psychology experiments revealed the powerful influence of authority figures on individuals' behavior, scammers posing as law enforcement leverage this authority dynamic to manipulate and deceive their victims. In the Milgram experiment, participants were instructed by an authority figure to administer electric shocks to another person, despite their personal moral beliefs. The key finding was that a significant number of participants were willing to obey the authority figure's commands, even when it involved potentially harmful actions. This demonstrated the extent to which individuals are inclined to comply with authority figures, even when it goes against their own ethical principles. Similarly, when scammers impersonate law enforcement officials, they exploit the inherent trust and obedience that individuals often have towards figures of authority. By assuming the guise of law enforcement, scammers create a sense of urgency and legitimacy, coercing their victims into complying with their demands. Victims may feel compelled to follow the instructions of the supposed law enforcement officer, believing that they are obligated to cooperate with an authority figure. The parallels between the Milgram experiment and scammers posing as law enforcement underscore the psychological mechanisms at play when individuals are confronted with authority figures. The influence of authority can override critical thinking and ethical considerations, leading individuals to unquestioningly follow commands, even in situations that may seem suspicious or unethical. By understanding how scammers exploit the dynamics of obedience to authority, individuals can become more vigilant and skeptical when faced with demands from supposed law enforcement officials. Awareness of these tactics can empower individuals to question the legitimacy of requests, verify the identity of individuals claiming to be in positions of authority, and protect themselves from falling victim to fraudulent schemes that prey on obedience to authority.

**Conclusion**

In summary, the Grandparent scam typically progresses through three key phases: contact, convince, and cash out.

1. Scammer contacts victim pretending to be:
    a. Grandchild in trouble
    b. Lawyer helping relative.
    c. Loved one being held hostage.
    d. Law enforcement

This deceptive tactic aims to evoke emotions of urgency and concern in the victim, compelling them to act quickly without verifying the authenticity of the situation.

2. Scammer isolates victim by:
    a. Begging victim not to inform parents.
    b. Threatening harm to loved ones if victim tells anyone.
    c. Scaring victim into believing that anyone they tell will be complicit in a crime and therefore the victim must act swiftly and discretely.

These manipulative strategies are designed to prevent the victim from seeking help or advice from others, thus increasing the chances of the scam's success.

3. Scammer(s) are not only the original caller but can include:
    a. Complicit courier
    b. Another scammer who steals driver accounts or rents driver accounts to ineligible drivers using stolen identities
    c. Bank or payment app employees.
    d. Telecom employees
    e. Relatives of the victim

This network of collaborators contributes to the complexity and sophistication of the scam, making it challenging for authorities to trace and apprehend all involved parties.

Pearl S. Buck, Pulitzer and Nobel prize winner, wrote "Our society must make it right and possible for old people not to fear the young or be deserted by them, for the test of a civilization is the way that it cares for its helpless members." The prevalence of the Grandparent Scam, alongside alarming trends such as child identity theft and online groups targeting minors to document self-harm, serves as indicators of the societal challenges we currently face. These issues reflect broader concerns about ethics, empathy, and the protection of individuals from malicious actors seeking to exploit their vulnerabilities. As such, it is imperative for readers to critically evaluate how our civilization responds to these threats and safeguards its most defenseless members. The impact of these fraudulent schemes extends beyond financial losses, encompassing emotional distress, loss of trust, and a sense of violation among victims. Therefore, a comprehensive approach that combines public awareness, regulatory measures, and community support is crucial in combating such fraudulent activities and upholding the values of a compassionate and just society.